\documentclass[aps,pra,twocolumn,superscriptaddress,showpacs,amsmath]{revtex4}
\usepackage{graphicx}
\usepackage{amssymb}

\newcommand{\ket}[1]{\, | #1 \rangle}

\newcommand{\sgn}{\mathrm{sgn}}

\newcommand{\hn}{\hat{n}}
\newcommand{\iim}{\mathrm{i}}
\newcommand{\be}{\begin{equation}}
\newcommand{\ee}{\end{equation}}
\newcommand{\bea}{\begin{eqnarray}}
\newcommand{\eea}{\end{eqnarray}}
\newcommand{\besa}{\begin{subeqnarray}}
\newcommand{\eesa}{\end{subeqnarray}}
\newcommand{\bean}{\begin{eqnarray*}}
\newcommand{\eean}{\end{eqnarray*}}
\newcommand{\hc}{\hat{c}}
\newcommand{\hcd}{\hat{c}^{\dagger}}

\begin{document}

\title{Frustrated collisions and unconventional pairing on a quantum superlattice}

\author{Manuel Valiente}
\affiliation{Institute of Electronic Structure and Laser, FORTH,
71110 Heraklion, Crete, Greece}
\affiliation{Institut f\"ur Physik,
Humboldt-Universit\"at zu Berlin, D-12489 Berlin, Germany}
\author{Matthias K\"uster}
\affiliation{Institut f\"ur Physik,
Humboldt-Universit\"at zu Berlin, D-12489 Berlin, Germany}
\author{Alejandro Saenz}
\affiliation{Institut f\"ur Physik,
Humboldt-Universit\"at zu Berlin, D-12489 Berlin, Germany}
\date{\today}

\begin{abstract}
We solve the problem of scattering and binding of two spin-$1/2$ fermions on a
one-dimensional superlattice with a period of twice the lattice spacing
analytically. We find the exact bound states and the scattering states, consisting of
a generalized Bethe ansatz augmented with an extra scattering product due to
"asymptotic" degeneracy. If a Bloch band is doubly occupied, the extra wave
can be a bound state in the continuum corresponding to a single-particle
interband transition. In all other cases, it corresponds to a quasi-momentum
changing, frustrated collision. 

\end{abstract}


\maketitle

\paragraph{Introduction.}

The study of electrons in discretized models with a periodicity larger than the
lattice spacing has a broad range of physical
applications. Among them, a superlattice model for ferroelectric perovskites
has been proposed \cite{Egami}, and its rich phase diagram qualitatively
explained \cite{Fabrizio}. The recent experimental advances with ultracold
bosonic or fermionic atoms loaded in optical lattices provide a neat
realization of superlattice Hamiltonians \cite{ReviewBloch}. These may allow
for experimental observation of fermionic d-wave superfluidity \cite{Rey} and magnetic phases
of cold spinor gases \cite{Rey2}. Moreover, one-dimensional electron systems can 
exhibit dimerization due to Peierls' instability \cite{Peierls}, thus
enlarging the periodicity of the system, by a factor of two, exactly
\cite{Lieb}. There is growing interest in the theoretical and numerical understanding of
the quantum phases of superlattice and dimerized models, as attested by the
attention received in the recent literature \cite{Rigol,Refs815}. Although
essentially exact numerical calculations in one
dimension can be performed, much of the physics is hidden in the numerical
data, and no unambiguous physical interpretation can be given in this way. In
particular, the exact origin of degeneracy in dimerized models is not known,
and may only be determined via an exact analytical solution. 

In this Communication, we consider the problem of two-electron
scattering and binding on a one-dimensional superlattice with
a period of twice the lattice spacing. After introducing two different
Hubbard models with non-trivial periodicity, we show that the Schr\"odinger
equation can be solved 
exactly. To do so, we generalize the two-body Bethe
ansatz \cite{Bethe} of the Hubbard model \cite{LiebWu} in a non-trivial manner, and center our discussion on one of the models, for concreteness. We find that, surprisingly, the collision of two electrons with
opposite spins either is frustrated --- the scattering states are degenerate --- or necessarily a bound state in the
continuum is produced upon collision. We also find all (up to four) bound states of the
system analytically. Our results constitute the first
quantum scattering problem on a non-trivial periodic potential, and with both
particles being mobile, to be solved analytically, and are also relevant to
the study of dilute quantum gases in optical superlattices where two-body
processes are dominant.   

\paragraph{Periodic Hubbard models.}

We start with the following general Hubbard Hamiltonian for spin-$1/2$
fermions 
\be
\hat{\mathcal{H}} = \sum_{j,\sigma} [ - J_j (\hcd_{j\sigma}\hc_{j+1\sigma} +
\mathrm{H.c.}) + \lambda_j \hn_{j\sigma}]+U\sum_j \hn_{j\uparrow}\hn_{j\downarrow}, \label{generalHam}
\ee
where $\hcd_{j\sigma}$ ($\hc_{j\sigma}$) is the creation (annihilation) operator of a fermion with
spin $\sigma=\uparrow,\downarrow$ at site $j$,
$\hn_{j\sigma}=\hcd_{j\sigma}\hc_{j\sigma}$ is the number operator, $J_j$
is the tunneling rate between adjacent sites $j$ and $j+1$, $\lambda_j$ is a
single-particle potential and $U$ is the on-site interaction between fermions
with opposite spin. We discuss here the case of an infinitely long lattice;
the finite lattice case with periodic boundary conditions (PBC) will be considered elsewhere \cite{preparation}.

There are two similar but different models which leave Hamiltonian
$\hat{\mathcal{H}}$ invariant under a translation $j\to j+\tau$ ($\tau \in
\mathbb{Z}$). The first is closely related to the so-called Peierls'
instability \cite{Lieb}, having a $\tau$-periodic tunneling rate $J_j = t +
\delta \cos(2\pi j /\tau)$, while its single-particle potential $\lambda_j$ vanishes. The second, most
commonly used, is that of a superlattice, with a constant tunneling $J_j = J$ $(>0)$
and a periodic potential $\lambda_j = \lambda \cos(2\pi j / \tau)$. We will
refer to the first of these models as Peierls-Hubbard (PHM), and to the latter
as ionic Hubbard (IHM).

The energy bands are obtained by substituting $\ket{\Psi_{\sigma}^s} = \sum_j \phi_{k,s} (j)e^{\iim k j}
\hcd_{j\sigma} \ket{0}$, with $\phi_{k,s}(j+\tau)=\phi_{k,s}(j)$, into the
Schr\"odinger equation  $\hat{\mathcal{H}}\ket{\Psi_{\sigma}^s} = \mathcal{E}_s \ket{\Psi_{\sigma}^s}$. For the simplest
case of $\tau=2$, considered from now on, the two models have common energy
dispersions, namely $\mathcal{E}_s (k) = (-1)^s \sqrt{(2J\cos(k))^2+\lambda^2}$, 
with $s=1,2$ labeling the two energy bands, and where we have set
$\sqrt{t^2-\delta^2}=J$ and $\delta=\lambda/2$, provided $|\delta|<|t|$. The
Bloch functions $\phi_{k,s}$ are, on the other hand, different for the two
models. Indeed, for the PHM $\phi_{k,s}(1)/\phi_{k,s}(0) = e^{\iim
  \varphi_{k,s}}$, $\varphi_{k,s}\in (0,2\pi]$
--- there is a phase- but not a density-modulation --- while for the IHM we
have
\be
\frac{\phi_{k,s}(1)}{\phi_{k,s}(0)} = \frac{\lambda-\mathcal{E}_s(k)}{2J\cos(k)}=-\frac{2J\cos(k)}{\mathcal{E}_s(k)+\lambda},
\ee
which satisfies $|\phi_{k,s}(1)|\ne |\phi_{k,s}(0)|$, and can even
represent a particle occupying only one sublattice (odd or even $j$) for $\lambda/J\to \infty$ or
$k=\pi/2$ (quasi-momenta are defined only mod $\pi$). 

The energy dispersions of the PHM and IHM are identical and
both systems have on-site interaction. Thus, if we can analytically solve one of them, the other is solvable as well. In this Communication, we restrict our discussion to the IHM since, as the reader can readily check, all the formalism discussed
below for this model can be applied to the PHM. We leave the solution of the PHM and further details to a
forthcoming paper \cite{preparation}. 


\paragraph{Wave functions of the two-body problem.}

Consider two interacting fermions of different spin components $\sigma_1\ne
\sigma_2$ in the singlet spin state \cite{footnote2} described by the IHM
with period $\tau=2$. The stationary Schr\"odinger equation for the
two-fermion {\it spatial} wave function $\ket{\Psi}$ reads, in first quantization,
 \begin{align}
&-J\sum_{\mu=-1,1}\big[\Psi(j_1+\mu,j_2)+\Psi(j_1,j_2+\mu)\big] \label{recrel} \\
&+\big[\lambda((-1)^{j_1}+(-1)^{j_2})
+ U\delta_{j_1,j_2}-E\big]\Psi(j_1,j_2)=0,\nonumber
\end{align}
where the wave function is symmetric under the exchange of the spatial 
coordinates, $\Psi(j_1,j_2)=\Psi(j_2,j_1)$.

In the limit of weak on-site interaction ($U=0$), the solutions to
Eq. (\ref{recrel}) are of the form $\Psi=\Psi_0$, 
\be
\Psi_0(j_1,j_2)=\hat{O}_S(\phi_{k_1,s_1}(j_1)\phi_{k_2,s_2}(j_2)e^{\iim(k_1j_1+k_2j_2)}),\label{bosonic}
\ee
with
$\hat{O}_S$ the symmetrization operator \cite{footnote3}, and eigenenergies 
$E=E_{s_1,s_2;k_1,k_2}=\mathcal{E}_{s_1}(k_1)+\mathcal{E}_{s_2}(k_2)$.
For very strong interaction $|U|/J \to \infty$ the solutions correspond to
``fermionized'' wave functions \cite{Rigol},
$\Psi= \Psi_F$,
\begin{align}
\Psi_F(j_1,j_2)=&\hat{O}_A(\phi_{k_1,s_1}(j_1)\phi_{k_2,s_2}(j_2)e^{\iim(k_1j_1+k_2j_2)})\nonumber
\\
\times & \sgn(j_1-j_2),\label{fermionized}
\end{align}
with $\hat{O}_A$ the antisymmetrization operator \cite{footnote3} and eigenenergies
given by $E_{s_1,s_2;k_1,k_2}$. Here we study the more interesting case of finite interaction
$0<|U|/J<\infty$, for which the analytic solutions were not known
previously. We discuss below all possible solutions corresponding to
scattering and bound states.

\paragraph{Scattering states.}
We first discuss the collisional states of two fermions. We invoke a periodically modulated generalization of the two-body Bethe ansatz, written in the
most convenient form for our purposes, defined for all $j_1$ and $j_2$ as
\be
\Psi_B(j_1,j_2)=\Psi_0(j_1,j_2)+B\Psi_F(j_1,j_2),\label{Bethe}
\ee
with $\Psi_0$ and $\Psi_F$ given by Eqs. (\ref{bosonic}) and
(\ref{fermionized}), and $B$ a $c$-number. The Bethe ansatz (\ref{Bethe}) is
an exact eigenfunction, i.e. it satisfies Eq. (\ref{recrel}), if the lattice is
homogeneous ($\lambda=0$). If $\lambda$ is finite, however, for the wave function to satisfy (\ref{recrel}) at
all $j_1,j_2$, we need to add an extra wave $\Psi_1$, with its corresponding constant
$C$, to the Bethe ansatz. The total wave function reads
\be
\Psi(j_1,j_2) = \Psi_B(j_1,j_2)+C\Psi_1(j_1,j_2)\label{wavefunction}.
\ee
Clearly, $\Psi_1$ and $\Psi_B$ must be, asymptotically ($|j_1-j_2|\ge 1$),
solutions to the Schr\"odinger equation (\ref{recrel}) with the same energy
and total quasi-momentum $K=k_1+k_2$, and
$\Psi_1/\Psi_B\ne \mathrm{const.}$ for all $B\in \mathbb{C}$. We note that $\Psi_B$ is a linearly
growing function of $|j_1-j_2|$ for vanishing relative quasi-momentum,
$k_1=k_2$ \cite{MV}; in
such case a generalized $K$-dependent scattering length can be {\it defined}
as $a\propto B^{-1}$
\cite{MVDPNygaard}; the divergence of the scattering length denotes a
bound state entering or exiting the continuum. The cases of total quasi-momentum $K=0,\pi/2$ also
have to be handled with special care by taking the limits $K\to 0,\pi/2$ {\it
  before} using the ansatz (\ref{wavefunction}) in Eq. (\ref{recrel}).

The set of band indices and quasi-momenta $\{s_1,s_2;k_1,k_2\}$ for the Bethe ansatz have the
corresponding energy $E_{s_1,s_2;k_1,k_2}$. Therefore, one has to find $\{s_1',s_2';k_1',k_2'\}$ so that the following two sets of conditions are fulfilled: (1i)
$E_{s_1,s_2;k_1,k_2}=E_{s_1',s_2';k_1',k_2'}$; (1ii) $k_1'+k_2'=k_1+k_2$; (2i) if $s_1=s_2=s$, either $s_1'\ne s_2'$ or $s_1'=s_2'=s$ and
$k_1'\ne k_1,k_2$; (2ii) if $s_1\ne s_2$, then $s_1'\ne s_2'$ and
$k_1'\ne k_1,k_2$. Once the
primed set is found one introduces the wave function (\ref{wavefunction}) in Eq. (\ref{recrel}) at $j_1=j_2$ even and odd, thus obtaining a set of two
linear equations in $B$ and $C$ that can be solved analytically, giving in turn the total wave function
$\Psi$. 

Since we are working
with a two-band model, and the single-particle bands are mirror symmetric,
$\mathcal{E}_{s}(k) = -\mathcal{E}_{s\pm 1} (k)$, we have to consider two
different possibilities, namely intra- and interband scattering. These cases
are qualitatively different from each other, as we shall see, and are therefore
considered separately.

\begin{figure}[t]
\includegraphics[width=0.38\textwidth]{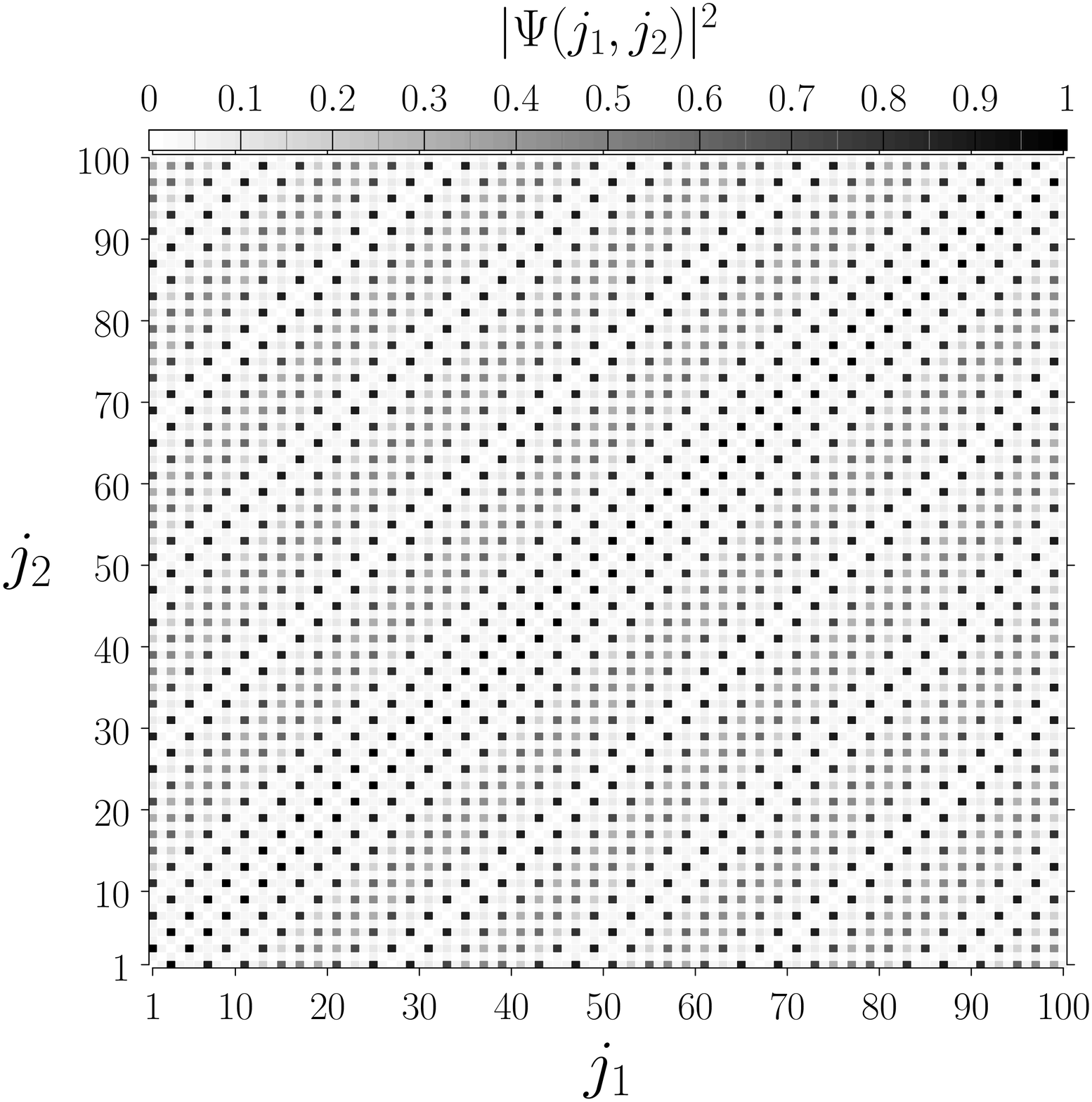}
\caption{Density $|\Psi|^2$ for scattering state with $s_1=s_2=1$, $k_1=0$,
  $k_2=0.45\pi$. $\Psi_1$ corresponds to a bound state in the
  continuum, Eq. (\ref{psiboundcont}); its contribution is large for small
  interparticle distances (around the diagonal). In this case $q=-0.275\pi + \iim
  0.671165$, $|B|=1.915167$ and $|C|=9.719295$ (see text). Values are
  normalized to the peak.}
\label{fig:bscont}
\end{figure}

(i) Intraband scattering. We assume that $\Psi_0$, interpreted as
incident wave, Eq. (\ref{bosonic}),
consists of two fermions with (real) quasi-momenta $k_1$ and $k_2$, occupying
the same band, $s_1=s_2=s$. As discussed in the previous
paragraph, we have to find a non-trivial set $\{s_1',s_2';k_1',k_2'\}$
corresponding to the same energy and total quasi-momentum $K$ as the incident state. In the majority of the
cases, the solution is given by a pair of complex quasi-momenta $k_1'\equiv q
= (K+\pi)/2 + \iim v$ and $k_2'=q^*$ (recall $K$ is defined mod $\pi$), with associated band indices $s_1'=s$ and $s_2'=s\pm
1$. Such solutions are outgoing bound states in the continuum, and represent an interband
transition ($s_1'\ne s_2'$). Note that interband transitions with real
quasi-momenta cannot happen since we have neither external forces nor energy
dissipation mechanisms. The only physically acceptable (exponentially
decaying) extra wave with complex quasi-momenta is given by $\Psi_1\equiv
\Psi_{\mathrm{bs}}^{q,q^*}$, with
\begin{align}
\Psi_{\mathrm{bs}}^{q_1,q_2^*}(j_1,j_2) &=
\big[\theta(j_1-j_2) \phi_{q_1,s_1'}(j_1)\phi_{q_2^*,s_2'}(j_2)e^{\iim (\ell_1 j_1+\ell_2j_2)}\nonumber \\
                &+ \tilde{\theta}(j_2-j_1)
                \phi_{q_1,s_1'}(j_2)\phi_{q_2^*,s_2'}(j_1)e^{\iim (\ell_1j_2+\ell_2j_1)}\big]\nonumber \\
                &\times e^{-|v||j_2-j_1|}, \label{psiboundcont}
\end{align} 
where, in general, $q_n\equiv\ell_n+\iim v$, and $\theta$ ($\tilde{\theta}$) is the step function being zero (one) at
$j_1-j_2=0$. In Fig. \ref{fig:bscont} we plot the density $|\Psi|^2$ for a
significant case ($\lambda/J=U/J=2$) in which all the parameters of the
Hamiltonian are in competition. As observed, the contribution of the
bound state in the continuum (\ref{psiboundcont}) is appreciable at small
interparticle distances, while the wave function behaves, for $|j_1-j_2|\gg
1$, as the periodically modulated Bethe ansatz $\Psi_B$.

This surprising result implies that paired states can be created upon two-body
collisions on a
superlattice even without external forces, phonons or dissipation, and are due
to the combination of the band structure and finite interactions. These states
should play a non-negligible role in the many-electron problem; we have
verified that the ground state of the system, with PBC, at
half-filling (4 lattice sites) and quarter-filling (8 sites) is always
a partially paired state for any non-zero $\lambda$ and $U>0$
\cite{preparation}. 

\begin{figure}[t]
\includegraphics[width=0.38\textwidth]{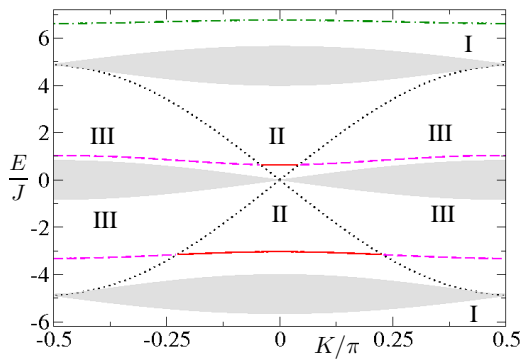}
\caption{(Color online) Two-body spectrum for $\lambda/J=U/J=2$. Shaded regions represent the
  two-body continua. Black dotted lines denote
  the boundaries between bound states of type II and III (see text). Red solid and magenta
  dashed lines
  are, respectively, energies of states of type II and III. Green
  dashed-dotted lines are energies of bound states of type I.}
\label{fig:spectrum}
\end{figure}

There are, on the other hand, cases for which the extra wave $\Psi_1$ has real quasi-momenta $k_1'$ and
$k_2'$. It has now the simple form
\be
\Psi_1(j_1,j_2)=\hat{O}_S(\phi_{k_1',s_1'}(j_1)\phi_{k_2',s_2'}(j_2)e^{\iim(k_1'j_1+k_2'j_2)}).\label{extrareal}
\ee
By energy conservation, it is obvious that the associated band
indices correspond to $s_1'=s_2'=s$. The energies and total quasi-momenta for which bound states in the
continuum or scattered extra waves are needed can be inferred from
Fig. \ref{fig:spectrum}, where the spectrum is plotted for 
$\lambda/J=U/J=2$. In the figure, the regions of the $s=2$ ($s=1$) continuum with
energies below (above) the dotted line (see the discussion on bound states below), correspond to cases for
which the extra wave has real quasi-momenta. These regions are already very
small -- but dense -- for the value of $\lambda$ under consideration. We note that as $\lambda/J$ gets larger, it becomes less likely that
$\Psi_1$ corresponds to a scattered wave. Since, for these cases, we can construct two different
wave functions (with $\Psi_F$ corresponding to $k_1,k_2$ or $k_1',k_2'$) or any
superposition thereof having the same
total quasi-momentum and energy, these states are degenerate (frustrated).  

(ii) Interband scattering. The two incident fermions, with quasi-momenta $k_1$
and $k_2$ are now in different bands, $s_1\ne s_2$. Energy conservation
implies that $\Psi_1$ has band indices $s_1'\ne s_2'$, and the non-trivial outgoing quasi-momenta
$k_1'$, $k_2'$ are always real -- there is no bound state in the
continuum -- and as a consequence interband scattering states are always
degenerate. The extra wave has again the simple form of Eq. (\ref{extrareal}),
and the total wave function $\Psi$, Eq. (\ref{wavefunction}), is calculated by
substitution in the Schr\"odinger equation at $j_1=j_2$, as we have already
explained. 

\paragraph{Bound states.} 

The system under consideration also supports bound states -- exponentially
decaying wave functions in the relative coordinate $|j_1-j_2|$ with energies
outside the continuum for each total quasi-momentum $K$ -- for both attractive and
repulsive on-site interaction (if $U>0$, these are sometimes called antibound
states \cite{Lorenzana}). Due to the periodicity of the Hamiltonian, we need,
as for the case of scattering states, two constants so that the Schr\"odinger
equation (\ref{recrel}) can be satisfied for all $j_1,j_2$. We propose the
following ansatz for the wave function
\be
\Psi(j_1,j_2) =
\Psi_{\mathrm{bs}}^{q_1,q_2^*}(j_1,j_2)+B\Psi_{\mathrm{bs}}^{\tilde{q}_1,\tilde{q}_2^*}(j_1,j_2),\label{boundstate}
\ee
with $q_n=\ell_n+\iim v$, $\tilde{q}_n=\tilde{\ell}_n+\iim\tilde{v}$, 
total quasi-momentum $K=\ell_1+\ell_2=\tilde{\ell}_1+\tilde{\ell}_2$ \cite{footnotepihalf}, and
$\Psi_{\mathrm{bs}}$ given by Eq. (\ref{psiboundcont}). The two wave functions
in Eq. (\ref{boundstate}) must have, asymptotically, the same energies, from
what we get a relation between $q_n$ and $\tilde{q}_n$; necessarily, the band
indices of the functions $\Psi_{\mathrm{bs}}$ are unequal (except for $K=0$
for which one of the functions $\Psi_{\mathrm{bs}}$ has $s_1=s_2$, $|v|<\infty$, and the other
has $s_1\ne s_2$ with $v=\infty$). The constant $B$
is then calculated for $j_1=j_2=j$ even (or odd), and $q_n$, $\tilde{q}_n$ are
varied self-consistently until the Schr\"odinger equation (\ref{recrel}) is
satisfied for $j$ even and odd, which yields the desired bound state
energy. Although it is possible to solve this problem for general $q_n$ and
$\tilde{q}_n$, it is simpler to distinguish all possible cases for
which Eq. (\ref{boundstate}) is a solution to Eq. (\ref{recrel}). There are three different types of bound states, based on the
combinations of $q_1$, $\tilde{q}_1$, as explained below.

Type I: bound states with energies below the $s_1=s_2=1$ or above the
$s_1=s_2=2$ continuum. These correspond to $q_1=K/2+\iim v$ and
$\tilde{q}_1=(K+\pi)/2+\iim \tilde{v}$, with $v\ne \tilde{v}$. Therefore, the
real parts of the quasi-momenta are fixed while $v$ and $\tilde{v}$ are 
calculated self-consistently with $B$.

Type II: energies in one of the band gaps and total
quasi-momenta $K\in (-\varphi,\varphi]$, with $0<\varphi<\pi/2$ depending on
the value of $U/J$ and $\lambda/J$. These have
$q_1=(K+\pi)/2+\iim v$ and $\tilde{q}_1=(K+\pi)/2+\iim \tilde{v}$, with $v\ne
\tilde{v}$, and their calculation is analogous to that of type I.

Type III: energies in one of the band gaps and total quasi-momenta $K\in
(-\pi/2,-\varphi]\cup (\varphi,\pi/2)$, with $0<\varphi<\pi/2$ being the same
number as for type II if bound states of that type exist in their range of
quasi-momenta. These have $q_1=\ell+\iim v$ and $\tilde{q}_1 =
\ell-\iim v$. One has to calculate, for a given starting value of $\ell$, the
relation between $v$ and $\ell$ so that the energy is real, and iterate
self-consistently until the Schr\"odinger equation is satisfied. 

All bound states for $\lambda/J=U/J=2$ are plotted in Fig. \ref{fig:spectrum},
where we clearly identify three different bound state bands. The first one
corresponds to a pair bound from the lowest scattering band with
$s_1=s_2=1$, and lies in the first band gap. The second band of bound states
is in the second band gap, and binding energies are evidently smaller than
those for the
first bound band, since they are bound from the $s_1\ne s_2$ band: the density
modulation prevents high occupancy of two particles at the same lattice site
if they are in different bands. The third band, above the continuum with
$s_1=s_2=2$, is similar (although not equivalent) to the first bound state
band. The dotted line in Fig. \ref{fig:spectrum} is the minimum (in absolute value) of
the energy $E_{1,2,q,q^*}$ with $q=(K+\pi)/2+\iim v$ for each value of the
total quasi-momentum $K$, and therefore denotes the boundaries $\pm \varphi$ between bound
states of type II and III. We note that, for larger values of $U/J$ (not
shown), there can be up to four bound states at certain values of the total quasi-momentum.


\paragraph{Conclusions.}

We have found that, on a one-dimensional superlattice, the non-trivial
underlying periodicity has important implications for the two-body
problem. We have derived exact, analytical solutions for the wave functions
which show, unambiguously, that partial pairing of fermions after their
scattering occurs over a
large range of parameters of the model studied, and corresponds to hitherto
unexplored bound states in the continuum. These states appear only if the
incident particles occupy the same band. In the case that the two fermions occupy
different bands, collisions produce a phase shift, but
also a second outgoing wave corresponding to quasi-momenta being different
from the incident ones. This implies that interband scattering is frustrated,
and the spectrum is degenerate. The implications of our results should persist in
the solution of the many-electron problem at non-zero densities, as can be
inferred from an analytical two-particle analysis in a finite lattice with
periodic boundary conditions, which involves a minor generalization of our
solutions \cite{preparation}. Our exact results are also of relevance to ultracold,
low-density gases in optical superlattices where the physics is most
influenced by pairwise collisions. Moreover, for bosons, our wave functions can be used to
construct trial functions of product (Jastrow) type for the many-body problem \cite{Jastrow}
in the dilute regime.      

\begin{acknowledgments}
We thank F. Cordob\'es-Aguilar for producing Fig. \ref{fig:bscont}. M.V. is
gratefully indebted to D. Petrosyan for useful discussions and comments. 
This work was supported by the EU network EMALI and the {\it Fonds der chemischen Industrie}.
\end{acknowledgments}


\begin{thebibliography}{99}
\bibitem{footnote2}
This is equivalent to a pair in the Bose-Hubbard model; triplet states and hard-core bosons are trivial.
\bibitem{footnotepihalf}
If $K=\pi/2$ there is no non-trivial ($\Psi \ne 0$) bound state.
\end{thebibliography}
\end{document}